\documentclass[12pt]{article}

\newcommand{\blind}{1}

\usepackage{authblk}
\usepackage{amsmath, amsthm, amssymb, bm}
\usepackage{graphicx, psfrag, epsf, subcaption}
\usepackage{enumerate}
\usepackage{natbib}
\usepackage{textcomp}
\usepackage[hyphens]{url} 
\usepackage{verbatim}

\usepackage{subfiles} 
\usepackage{dsfont}
\usepackage{xcolor} 
\usepackage{hyperref}
\hypersetup{%
	colorlinks = false
}

\usepackage{verbatim}

\usepackage[utf8]{inputenc}

\usepackage{float} 

\usepackage{listings}
\usepackage{adjustbox}
\usepackage{multirow} 

\usepackage{lineno} 

\usepackage{moreverb}
\usepackage{amsmath}
\usepackage{amsthm}
\usepackage{array}
\newcolumntype{P}[1]{>{\centering\arraybackslash}p{#1}}
\newcolumntype{M}[1]{>{\centering\arraybackslash}m{#1}}
\usepackage{dsfont}
\usepackage{threeparttable}
\usepackage{booktabs}
\usepackage{rotating}
\usepackage{hyperref}
\hypersetup{hidelinks}
\usepackage{citeall}
\usepackage{float}
\usepackage{mathtools}
\usepackage{xcolor}
\usepackage[title]{appendix}

\usepackage{lineno}

\newcommand\BibTeX{{\rmfamily B\kern-.05em \textsc{i\kern-.025em b}\kern-.08em
		T\kern-.1667em\lower.7ex\hbox{E}\kern-.125emX}}

\usepackage[ruled,vlined]{algorithm2e}

\begin{document}
	
	\def\spacingset#1{\renewcommand{\baselinestretch}%
		{#1}\small\normalsize} \spacingset{1}

	\if1\blind
	{
		\title{\bf Dynamic Treatment Regimes with Replicated Observations Available for Error-prone Covariates: a Q-learning Approach}
		
		\author[1]{Dan Liu\thanks{Corresponding author, email: dliu372@uwo.ca}}
		\author[1,2]{Wenqing He}
		\affil[1]{Department of Statistical and Actuarial Sciences, University of Western Ontario, London, Ontario, Canada N6A 5B7}
		\affil[2]{Department of Oncology, University of Western Ontario, London, Ontario, Canada N6A 5W9}
		\date{}                     
		\setcounter{Maxaffil}{0}
		\renewcommand\Affilfont{\itshape\small}
		\maketitle
	} \fi
	
	\if0\blind
	{
		\title{ \bf  Dynamic Treatment Regimes with Replicated Observations Available for Error-prone Covariates: a Q-learning Approach}
		
		\author[1]{  }
		\setcounter{Maxaffil}{0}
		\renewcommand\Affilfont{\itshape\small}
		\maketitle
	} \fi
	
	\begin{abstract}
		Dynamic treatment regimes (DTRs) have received an increasing interest in recent years. DTRs are sequences of treatment decision rules tailored to patient-level information. The main goal of the DTR study is to identify an optimal DTR, a sequence of treatment decision rules that yields the best expected clinical outcome. Q-learning has been considered as one of the most popular regression-based methods to estimate the optimal DTR. However, it is rarely studied in an error-prone setting, where the patient information is contaminated with measurement error. In this paper, we study the effect of covariate measurement error on Q-learning and propose a correction method to correct the measurement error in Q-learning. Simulation studies are conducted to assess the performance of the proposed method in Q-learning. We illustrate the use of the proposed method in an application to the sequenced treatment alternatives to relieve depression data.
	\end{abstract}
	
	\noindent%
	{\it Keywords:} Covariate measurement error, Q-learning, regression calibration, replicate data
	
	\spacingset{1.45} 
	
	
	\section{Introduction}\label{intro}

	Precision medicine is a novel approach in healthcare in which the treatment decisions are tailored to patient-specific information. It is particularly beneficial for chronic diseases that involve multiple treatment stages. For example, in the Sequenced Treatment Alternatives to Relieve Depression (STAR*D) study, the medical care went through more than two stages of intervention with an assessment of the depression severity for patients using a Quick Inventory of Depressive Symptomatology (QIDS) score \citep{rush200316}. One objective of the study is to identify a sequence of treatment decision rules across multiple stages that minimizes the severity of depression based on the relevant patient's characteristics.  
	
	To achieve the goal of identifying the optimal sequential treatments throughout the period, the concept of dynamic treatment regimes is introduced. A dynamic treatment regime is defined as a sequence of treatment decision rules by taking a patient's characteristics and treatment history as input \citep{chakraborty2013statistical}. The statistical goal is to determine an optimal DTR, an optimal sequence of treatment decision rules that leads to the best expected outcome.   
	
	Many statistical methods have been developed in the past to estimate an optimal DTR with a continuous outcome. Classical regression-based methods that indirectly estimate an DTR include Q-learning \citep{watkins1989learning, chakraborty2014dynamic}, G-estimation \citep{robins2004optimal}, A-learning \citep{murphy2003optimal, schulte2014q}, and regret-regression \citep{henderson2010regret}. Bayesian approaches was also established \citep{thall2000evaluating, arjas2010optimal, murray2018bayesian,  yu2023bayesian}. \cite{wallace2015doubly} proposed dynamic weighted ordinary least squares (dWOLS), a doubly robust estimation method, which combines the implementation simplicity of Q-learning and the double robustness property of G-estimation. Q-learning has been in particular widely studied with estimation and inference in various contexts \citep{chakraborty2010inference, qian2011performance, moodie2012q, nahum2012q, laber2014interactive, zhu2019proper,  ertefaie2021robust}. This approach is appealing and favorable due to its simplicity to understand and implement and resulted consistent estimators when the model is correctly specified \citep{chakraborty2013statistical}. 
	
	Despite its popularity and advantages, Q-learning, like other existing methods, assumes that the variables used in the modeling process are measured accurately, which may not be a realistic assumption in many circumstances. In the STAR*D study, the QIDS score was reported by both clinicians and patients, which may deviate from the actual severity level of depression. Consequently, the subjective reported QIDS scores may not be the true QIDS and are subject to measurement error. It is concerning that using the reported QIDS scores without error correction in the analysis may lead to misleading results. Therefore, it is imperative to study the effect of measurement error in patient characteristics on DTR when repeated measurements are available, where the reported QIDS scores of both the clinicians and the patients are viewed as the repeated measurements of the true QIDS score.    
	
	The effects of covariate measurement error have been extensively studied in the literature. Severe consequences of ignoring the measurement error in the parameter estimation were observed and theoretically justified \citep{carroll2006measurement, yi2017statistical}. Methods such as regression calibration have been used to correct the measurement error when validation data or replicated data are available, and it performs well in linear models \citep{carroll2006measurement}. A simulation-extrapolation method was developed as an alternative to the regression calibration method, but it suffers high computational intensity \citep{stefanski1995simulation}. Maximum likelihood method was utilized in linear and logistic models to correct the measurement error \citep{bartlett2009linear}. Multiple imputation-related methods have also been proposed for repeated measurements \citep{keogh2014toolkit, gray2018use}.  
	
	To our knowledge, dynamic treatment regimes and measurement error have been richly but separately studied in the literature. The study of DTR in the context of measurement error is scarcely considered except on  certain occasions. The dWOLS with covariates subject to measurement error was studied and corrected using the regression calibration method \citep{spicker2020measurement}. However, it remains unclear whether and how much the covariate measurement error affects the performance of Q-learning. In this paper, we aim to investigate the effect of measurement error in covariates on Q-learning when repeated measurements are available. 
	
	The rest of the article is structured as follows. In Section \ref{qlearn}, we present the Q-learning framework in the absence of mismeasured covariates. In Section \ref{qlearnME}, we describe the Q-learning with covariate measurement error and propose the use of the regression calibration method in Q-learning to correct the covariate measurement error in Section \ref{correction}. Simulation studies are carried out to examine the performance of the proposed method in Section \ref{sim}. In Section \ref{da}, we apply the proposed method to the STAR*D study. The findings and discussion are summarized in Section \ref{conclude}.

	\section{Q-learning without Measurement Error} \label{qlearn}
	
	\subsection{Notations and Model Framework}
	For simplicity, we restrict the notations and model framework to two decision points. The observed data trajectory follows \{$\boldsymbol{X_{1}}$, $\boldsymbol{Z_{1}}$, $A_{1}$, $\boldsymbol{X_{2}}$, $\boldsymbol{Z_{2}}$, $A_{2}$, $ Y $\}, where $\boldsymbol{X_{j}}$ and $\boldsymbol{Z_{j}}$ are error-prone covariate vector and error-free covariate vector ($ j $ = 1, 2), respectively. We consider a situation where the true covariate $\boldsymbol{X_{j}}$ is not observable at stage $ j $, but there are up to $k_{j}$  replicate surrogates observed for $\boldsymbol{W_{j}} = (\boldsymbol{W_{j1}}, ..., \boldsymbol{W_{jk_{j}}})$, where $\boldsymbol{W_{jl}}$ ($l$ = 1, ..., $ k_{j} $) denotes a surrogate or mismeasured version of $\boldsymbol{X_{j}}$. The classical additive model is assumed to accommodate the relationship between $\boldsymbol{W_{jl}}$ and $\boldsymbol{X_{j}}$, that is,  $\boldsymbol{W_{jl}}$ = $\boldsymbol{X_{j}}$ + $\boldsymbol{e_{jl}}$, where the $\boldsymbol{e_{jl}}$ follows a normal distribution with mean $\boldsymbol{0}$ and covariance $\boldsymbol{\Sigma_{ee}}$ and is independent of each other and of all other variables. The binary treatment $A_{j} \in \{0, 1\}$ is assigned at stage $ j $. $ Y $ is a continuous outcome observed at the end of the second stage. History $\boldsymbol{H_{j}}$ with realization $\boldsymbol{h_{j}}$ is defined as the accumulative information collected up to $ j $-th stage prior to making treatment decision $A_{j}$, $\boldsymbol{H_{1}} $ = ($\boldsymbol{X_{1}}$, $\boldsymbol{Z_{1}}$) and $ \boldsymbol{H_{2}}  $ = ($\boldsymbol{X_{1}}$, $\boldsymbol{Z_{1}}$, $A_{1}$, $\boldsymbol{X_{2}}$, $\boldsymbol{Z_{2}}$). A dynamic treatment regime $\boldsymbol{a} $ = \{$a_{1}$, $a_{2}$\} consists of treatment decision rules, where $a_{j}$ = $a_{j}(\boldsymbol{h_{j}})$ is the treatment received at stage $ j $. $ \boldsymbol{a^{opt}} $ denotes an optimal DTR such that $ \boldsymbol{a^{opt}} $ = \{$a^{opt}_{1}$, $a^{opt}_{2}$\}, where $a^{opt}_{j}$ = $a^{opt}_{j}(\boldsymbol{h_{j}})$ is the optimal treatment received at stage $ j $. In the Q-learning framework, two assumptions are made \citep{chakraborty2013statistical}: 
	
	(A1) $\textit{Stable unit treatment value}$: an individual's outcome is not influenced by other individuals' treatment allocation.   
	
	(A2) $\textit{No unmeasured confounders}$: for any possible treatment rule, the treatment $A_{j}$ received at stage $ j $ is independent of any future (potential) covariate or outcome, conditional on the history $\boldsymbol{H_{j}}$.   
	
	\subsection{Q-learning}
	
	Q-learning originates from reinforcement learning and has become one of the most popular regression-based methods to estimate an optimal DTR \citep{watkins1989learning, chakraborty2014dynamic, sutton2018reinforcement}. The Q-learning is modeled by stage-specific Q-functions, which measure the expected future outcome conditional on a patient's history and treatment action. In a two-stage setting, the Q-functions are defined as follows 
	\begin{equation} \label{qfun0}
		\begin{split}
			Q_{2}(\boldsymbol{H_{2}}, A_{2}) & = E[Y | \boldsymbol{H_{2}}, A_{2}],  \\
			Q_{1}(\boldsymbol{H_{1}}, A_{1}) & = E[\underset{a_{2}}{\max} Q_{2}(\boldsymbol{H_{2}}, a_{2}) | \boldsymbol{H_{1}}, A_{1}].    
		\end{split}
	\end{equation}
	
	The Q-functions are often unknown and need to be estimated using the data. A common approach is to parameterize $Q_{j}(\boldsymbol{H_{j}}, A_{j})$ at stage $  j $ via regression models
	\begin{equation} \label{qfun1}
		Q_{j}(\boldsymbol{H_{j}}, A_{j};\boldsymbol{\beta_{j}}, \boldsymbol{\psi_{j}}) = f(\boldsymbol{H_{j0}}; \boldsymbol{\beta_{j}}) + g(\boldsymbol{H_{j1}}, A_{j}; \boldsymbol{\psi_{j}}).
	\end{equation}
	In (\ref{qfun1}), $Q_{j}(\boldsymbol{H_{j}}, A_{j})$ is divided into a treatment-free component $f$($\boldsymbol{H_{j0}}; \boldsymbol{\beta_{j}}$) and a blip component $g(\boldsymbol{H_{j1}}, A_{j}; \boldsymbol{\psi_{j}}$), where $f$($\boldsymbol{H_{j0}}; \boldsymbol{\beta_{j}}$) is a function of $\boldsymbol{H_{j0}}$, and $g(\boldsymbol{H_{j1}}, A_{j}; \boldsymbol{\psi_{j}}$) is a function of both $\boldsymbol{H_{j1}}$ and $A_{j}$, with $\boldsymbol{H_{j0}}$ and $\boldsymbol{H_{j1}}$ being the subsets of $\boldsymbol{H_{j}}$. The covariates collected in $\boldsymbol{H_{j1}}$ are called tailoring variables. The functions $f(\cdot$) and $g(\cdot$) can be specified in any form, and a simple specification of the Q-functions is a linear form
	\begin{equation}\label{qfun2}
		Q_{j}(\boldsymbol{H_{j}}, A_{j};\boldsymbol{\beta_{j}}, \boldsymbol{\psi_{j}}) = \boldsymbol{\beta_{j}^{T}H_{j0}} +  \big(\boldsymbol{\psi_{j}^{T}H_{j1}}\big)A_{j}.
	\end{equation}
	
	Q-learning relies on a backward induction procedure for parameter estimation. In the set of parameters  ($\boldsymbol{\beta_{j}}, \boldsymbol{\psi_{j}}$), the interest focuses on blip parameter $\boldsymbol{\psi_{j}}$ in that it directly determines the optimal treatment. The treatment that maximizes the $Q_{j}(\boldsymbol{H_{j}}, A_{j};\boldsymbol{\hat{\beta}_{j}}, \boldsymbol{\hat{\psi}_{j}})$ can be expressed as
	\begin{align*}\label{aopt}
		\hat{a}_{j}^{opt} = \underset{a_{j}}{\arg\max} \ Q_{j}(\boldsymbol{h_{j}}, a_{j}; \boldsymbol{\hat{\beta}_{j}}, \boldsymbol{\hat{\psi}_{j}}).
	\end{align*}
	Given $A_{j} \in \{0, 1\}$ in (\ref{qfun2}), $\hat{a}_{j}^{opt}$ is further inferred to be $\hat{a}_{j}^{opt}$ = $\mathds{1}\big(\boldsymbol{\hat{\psi}_{j}^{T}h_{j1}} > 0\big)$, with $\mathds{1}(\cdot)$ being the indicator function. It implies that $\hat{a}_{j}^{opt}$ = 1 if $\boldsymbol{\hat{\psi}_{j}^{T}h_{j1}} > $ 0, and $\hat{a}_{j}^{opt}$ = 0 otherwise. 
	
	The estimator ($ \boldsymbol{\hat{\beta}_{2}}, \boldsymbol{\hat{\psi}_{2}} $) can be obtained by using the ordinary least squares (OLS) for $Q_{2}(\boldsymbol{H_{2}}, A_{2})$. Having worked backward recursively, the first stage parameter estimation depends on a pseudo-outcome $\widetilde{Y}_{1}$, which is defined as the future reward had the patients received the second stage optimal treatment, denoted as $\widetilde{Y}_{1} = \underset{a_{2}}{\max} Q_{2}(\boldsymbol{H_{2}},  a_{2}) $. Following the new expression for the estimated second stage optimal treatment $\hat{a}_{2}^{opt}$, the pseudo-outcome $\widetilde{Y}_{1}$ can be further written as
	\begin{equation} \label{pseudo1}
		\begin{split}
			\widetilde{Y}_{1} & = \underset{a_{2}}{\max} Q_{2}(\boldsymbol{H_{2}},  a_{2}; \boldsymbol{\hat{\beta}_{2}}, \boldsymbol{\hat{\psi}_{2}}) =  \boldsymbol{\hat{\beta}_{2}^{T}H_{20}} + (\boldsymbol{\hat{\psi}_{2}^{T}H_{21}})\mathds{1}\big(\boldsymbol{\hat{\psi}_{2}^{T}H_{21}} > 0\big).  
		\end{split}
	\end{equation}
	The estimation of parameters ($\boldsymbol{\beta_{1}}, \boldsymbol{\psi_{1}}$) can then be obtained by treating the pseudo-outcome $ \widetilde{Y}_{1} $ as the observed outcome for the first stage using the ordinary least squares.
	
	A two-stage linear Q-learning algorithm in DTR can be summarized in the following steps:  
	
	1. Parameterize the stage 2 Q-function
	\begin{center}
		$Q_{2}(\boldsymbol{H_{2}}, A_{2};\boldsymbol{\beta_{2}}, \boldsymbol{\psi_{2}})  = E[Y|\boldsymbol{H_{2}}, A_{2}] = \boldsymbol{\beta_{2}^{T}H_{20}} + \big(\boldsymbol{\psi_{2}^{T}H_{21}}\big)A_{2}$.   
	\end{center}
	
	2. Apply OLS procedure and obtain the stage 2 estimator $(\boldsymbol{\hat{\beta}_{2}},  \boldsymbol{\hat{\psi}_{2}})$
	\begin{center}
		$(\boldsymbol{\hat{\beta}_{2}},  \boldsymbol{\hat{\psi}_{2}})$ = $\underset{(\boldsymbol{\beta_{2}}, \boldsymbol{\psi_{2}})}{\arg\min}$ $\frac{1}{n}$ $\sum_{i=1}^{n}$ $\big(Y_{i} - Q_{2}(\boldsymbol{H_{2}}, A_{2};\boldsymbol{\beta_{2}}, \boldsymbol{\psi_{2}})\big)^{2}$.   
	\end{center}
	
	3. Derive the stage 2 optimal treatment as  $\hat{a}_{2}^{opt}$ = $\mathds{1}\big(\boldsymbol{\hat{\psi}_{2}^{T}h_{21}} > 0\big)$.    
	
	4. Construct the pseudo-outcome for estimation at stage 1 
	\begin{align*}
		\widetilde{Y}_{1} = \boldsymbol{\hat{\beta}_{2}^{T}H_{20}} + (\boldsymbol{\hat{\psi}_{2}^{T}H_{21}})\mathds{1}\big(\boldsymbol{\hat{\psi}_{2}^{T}H_{21}} > 0\big).
	\end{align*}
	
	5. Parameterize the stage 1 Q-function 
	\begin{center}
		$Q_{1}(\boldsymbol{H_{1}}, A_{1};\boldsymbol{\beta_{1}}, \boldsymbol{\psi_{1}})  = \boldsymbol{\beta_{1}^{T}H_{10}} + \big(\boldsymbol{\psi_{1}^{T}H_{11}}\big)A_{1}$.   
	\end{center}

	6. Apply OLS procedure and obtain the stage 1 estimator $(\boldsymbol{\hat{\beta}_{1}},  \boldsymbol{\hat{\psi}_{1}})$  
	\begin{center}
		$(\boldsymbol{\hat{\beta}_{1}}, \boldsymbol{\hat{\psi}_{1}})$ = $\underset{(\boldsymbol{\beta_{1}}, \boldsymbol{\psi_{1}})}{\arg\min}$ $\frac{1}{n}$ $\sum_{i=1}^{n}$ $\big(\widetilde{Y}_{i1} - Q_{1}(\boldsymbol{H_{1}}, A_{1};\boldsymbol{\beta_{1}}, \boldsymbol{\psi_{1}})\big)^{2}$.  
	\end{center}  
	
	7. Derive the stage 1 optimal treatment as  $\hat{a}_{1}^{opt}$ =  $\mathds{1}\big(\boldsymbol{\hat{\psi}_{1}^{T}h_{11}} > 0\big)$.  
	
	Q-learning enjoys the advantage of simplicity in implementation. By following the procedures above, the regression parameters can be consistently estimated if the outcome model is correctly specified \citep{chakraborty2013statistical}. 
	
	\section{Q-learning with Measurement Error} \label{qlearnME} 
	When the true covariate $ \boldsymbol{X_{j}}$ is absent but the replicate surrogates $\boldsymbol{W_{j}}$ are available at stage $ j $, the data trajectory is then replaced by 
	\begin{align*}
		\{\boldsymbol{W_{1}}, \boldsymbol{Z_{1}}, A_{1}, \boldsymbol{W_{2}}, \boldsymbol{Z_{2}}, A_{2}, Y\}.   
	\end{align*}
	
	In this case, the naive histories are formed as $\boldsymbol{H^{n}_{1}} $ = ($ \boldsymbol{W_{1}} $, $\boldsymbol{Z_{1}}$) and $ \boldsymbol{H^{n}_{2}}  $ = ($ \boldsymbol{W_{1}} $, $\boldsymbol{Z_{1}}$, $A_{1}$, $ \boldsymbol{W_{2}} $, $\boldsymbol{Z_{2}}$). As a result, the Q-functions that use the naive histories are called naive Q-functions, which treat the replicate surrogates as the observations of the true covariates. Then, the naive Q-functions are given by  
	\begin{align*}
		Q_{2}(\boldsymbol{H^{n}_{2}}, A_{2};\boldsymbol{\beta_{2}^{n}}, \boldsymbol{\psi_{2}^{n}})  & = f(\boldsymbol{H^{n}_{20}}; \boldsymbol{\beta_{2}^{n}}) + g(\boldsymbol{H^{n}_{21}}, A_{2}; \boldsymbol{\psi_{2}^{n}}),   \\
		Q_{1}(\boldsymbol{H^{n}_{1}}, A_{1};\boldsymbol{\beta_{1}^{n}}, \boldsymbol{\psi_{1}^{n}}) &  = f(\boldsymbol{H^{n}_{10}}; \boldsymbol{\beta_{1}^{n}}) + g(\boldsymbol{H^{n}_{11}}, A_{1}; \boldsymbol{\psi_{1}^{n}}).   
	\end{align*}
	If the functions $f(\cdot$) and $g(\cdot$) are modeled linearly, then the naive Q-function at stage $ j $ is given by
	\begin{equation} \label{qfun4}
		Q_{j}(\boldsymbol{H_{j}^{n}}, A_{j};\boldsymbol{\beta_{j}^{n}}, \boldsymbol{\psi_{j}^{n}}) = \boldsymbol{\beta_{j}^{nT}H^{n}_{j0}} + \big(\boldsymbol{\psi_{j}^{nT}H^{n}_{j1}}\big)A_{j}.   
	\end{equation}
	
	The naive Q-functions ($\ref{qfun4}$) are different from ($\ref{qfun2}$) in a sense that the original history is replaced with the naive history. By applying the OLS, the naive estimator ($\boldsymbol{\hat{\beta}^{n}_{j}}$, $\boldsymbol{\hat{\psi}^{n}_{j}}$) is obtained. It has been well documented that ignoring the measurement error may result in biased estimation \citep{carroll2006measurement, yi2017statistical}. Thus, it is reasonable to believe that the naive estimator ($\boldsymbol{\hat{\beta}^{n}_{j}}$, $\boldsymbol{\hat{\psi}^{n}_{j}}$) may be biased from ($\boldsymbol{\beta_{j}}$, $\boldsymbol{\psi_{j}}$). Since the blip parameter $ \boldsymbol{\psi} $ = ($ \boldsymbol{\psi_{2}} $, $ \boldsymbol{\psi_{1}} $) is of primary interest for estimation, and the naive blip estimator  $\boldsymbol{\hat{\psi}^{n}} $ = ($\boldsymbol{\hat{\psi}_{2}^{n}}$, $\boldsymbol{\hat{\psi}_{1}^{n}}$) may be biased from $ \boldsymbol{\psi} $, the estimated optimal treatment $\hat{a}_{j}^{opt}$ = $\mathds{1}\big(\boldsymbol{\hat{\psi}_{j}^{T}h_{j1}} > 0\big)$ may be wrong. We are motivated to correct the biases in the parameter estimation by applying an approximation $\boldsymbol{\hat{X}_{j}}$ to $\boldsymbol{X_{j}}$ using the available replicate surrogates in the data.

	\section{Regression Calibration} \label{correction}
	
	The regression calibration method is one of the methods used to address the covariate measurement error \citep{prentice1982covariate}. It has become a widely used error correction method when either a validation sample or replicated observations of the error-prone variables are available. In this paper, we focus on the study with replicate data of size $n$. The key idea of regression calibration is to find the estimates $\boldsymbol{\hat{X}}$ of $ \boldsymbol{X} $ using the available replicate surrogates $ \boldsymbol{W} $ and proceed with the analysis using the estimates $\boldsymbol{\hat{X}}$ so that the bias caused by the measurement error can be corrected.     
	
	In Q-learning, for any $i$-th patient ($i$ = 1, ..., $ n $) at stage $ j $, consistent regression calibration estimates $\boldsymbol{\hat{X}_{ij}}$ of $\boldsymbol{X_{ij}}$ can be obtained using the replicate surrogates $\boldsymbol{W_{ij}}$ \citep{carroll2006measurement}
	\begin{equation} \label{RCrep}
		\boldsymbol{\hat{X}_{ij}} = \hat{\mu}_{w} + 
		\left[\begin{array}{cc} 
			\hat{\Sigma}_{xx}, \ \hat{\Sigma}_{xz} 
		\end{array}\right]  
		\left[ \begin{array}{cc}
			\hat{\Sigma}_{xx} + \hat{\Sigma}_{ee}/k_{ij}   &  \hat{\Sigma}_{xz} \\
			\hat{\Sigma}_{xz}^{T}     &  \hat{\Sigma}_{zz}
		\end{array}    \right]^{-1}
		\left( \begin{array}{cc}
			\overline{\boldsymbol{W}}_{\boldsymbol{ij}} - \hat{\mu}_{w}    \\
			\boldsymbol{Z_{ij}} - \hat{\mu}_{z}
		\end{array}  \right),
	\end{equation}
	where
	\begin{align*}
		\overline{\boldsymbol{W}}_{\boldsymbol{ij}} & = \frac{1}{k_{ij}}\sum_{l=1}^{k_{ij}} \boldsymbol{W_{ijl}},   \\
		\hat{\mu}_{x} & = \hat{\mu}_{w} = \sum_{i=1}^{n}k_{ij}\overline{\boldsymbol{W}}_{\boldsymbol{ij}} \bigg/ \sum_{i=1}^{n}k_{ij}, \\
		\hat{\mu}_{z} & = \overline{\boldsymbol{Z}}_{j}, \\
		\nu & = \sum_{i=1}^{n}k_{ij} - \sum_{i=1}^{n}k_{ij}^{2} \bigg/ \sum_{i=1}^{n}k_{ij}, \\
		\hat{\Sigma}_{xx} & = \Bigg[\Bigg\{ \sum_{i=1}^{n}k_{ij}(\overline{\boldsymbol{W}}_{\boldsymbol{ij}} - \hat{\mu}_{w})(\overline{\boldsymbol{W}}_{\boldsymbol{ij}} - \hat{\mu}_{w})^{T} \Bigg\} - (n - 1)\hat{\Sigma}_{ee} \Bigg] \bigg/ \nu,   \\
		\hat{\Sigma}_{xz} & = \sum_{i=1}^{n}k_{ij}(\overline{\boldsymbol{W}}_{\boldsymbol{ij}} - \hat{\mu}_{w})(\boldsymbol{Z_{ij}} - \hat{\mu}_{z})^{T} \big/\nu,  \\
		\hat{\Sigma}_{zz} & = (n-1)^{-1} \sum_{i=1}^{n}(\boldsymbol{Z_{ij}} - \hat{\mu}_{z})(\boldsymbol{Z_{ij}} - \hat{\mu}_{z} )^{T}, \\
		\hat{\Sigma}_{ee} & = \sum_{i=1}^{n} \sum_{l=1}^{k_{ij}} (\boldsymbol{W_{ijl}} - \overline{\boldsymbol{W}}_{\boldsymbol{ij}})(\boldsymbol{W_{ijl}} - \overline{\boldsymbol{W}}_{\boldsymbol{ij}})^{T} \Big/ \sum_{i=1}^{n}(k_{ij} - 1).    
	\end{align*}
	
	Then, by replacing the unobserved $\boldsymbol{X_{j}}$ with the consistent estimates $\boldsymbol{\hat{X}_{j}}$, the data trajectory becomes
	\begin{align*}
		\{\boldsymbol{\hat{X}_{1}}, \boldsymbol{Z_{1}}, A_{1}, \boldsymbol{\hat{X}_{2}}, \boldsymbol{Z_{2}}, A_{2}, Y\}.  
	\end{align*}
	From the new data trajectory, we can obtain the regression calibration histories in a form $\boldsymbol{H_{1}^{rc}} $ = ($ \boldsymbol{\hat{X}_{1}} $, $\boldsymbol{Z_{1}}$) and $ \boldsymbol{H_{2}^{rc}}  $ = ($ \boldsymbol{\hat{X}_{1}} $, $\boldsymbol{Z_{1}}$, $A_{1}$, $\boldsymbol{\hat{X}_{2}} $, $\boldsymbol{Z_{2}}$). The Q-functions based on the regression calibration histories are given by
	\begin{equation}
		\begin{aligned} \label{qfun5}
			Q_{2}(\boldsymbol{H^{rc}_{2}}, A_{2};\boldsymbol{\beta^{rc}_{2}}, \boldsymbol{\psi^{rc}_{2}})  & = f(\boldsymbol{H^{rc}_{20}}; \boldsymbol{\beta_{2}^{rc}}) + g(\boldsymbol{H^{rc}_{21}}, A_{2}; \boldsymbol{\psi_{2}^{rc}}),   \\
			Q_{1}(\boldsymbol{H^{rc}_{1}}, A_{1};\boldsymbol{\beta^{rc}_{1}}, \boldsymbol{\psi^{rc}_{1}}) &  = f(\boldsymbol{H^{rc}_{10}}; \boldsymbol{\beta_{1}^{rc}}) + g(\boldsymbol{H^{rc}_{11}}, A_{1}; \boldsymbol{\psi_{1}^{rc}}).      
		\end{aligned}
	\end{equation}
	If each Q-function in (\ref{qfun5}) is modeled linearly, then it can be expressed as
	\begin{equation}\label{qfun6}
		Q_{j}(\boldsymbol{H_{j}^{rc}}, A_{j};\boldsymbol{\beta_{j}^{rc}}, \boldsymbol{\psi_{j}^{rc}}) = \boldsymbol{\beta_{j}^{rcT}H^{rc}_{j0}} + \big(\boldsymbol{\psi_{j}^{rcT}H^{rc}_{j1}}\big)A_{j}.   \\
	\end{equation}
	
	The modified Q-functions ($\ref{qfun5}$) and ($\ref{qfun6}$) are formalized based on the regression calibration histories, which consist of the corrected estimates for the error-prone covariates and other variables. Then, the estimator ($\boldsymbol{\hat{\beta}^{rc}_{j}}$, $\boldsymbol{\hat{\psi}^{rc}_{j}}$) obtained from the Q-functions ($\ref{qfun5}$) is the regression calibration estimator. It is discussed that the regression calibration method yields consistent estimators in linear models but is approximately consistent in nonlinear models \citep{carroll2006measurement, yi2017statistical}. That is, if the Q-function is in the form of (\ref{qfun6}), the regression calibration blip estimator $\boldsymbol{\hat{\psi}^{rc}_{j}}$ is a consistent estimator of $\boldsymbol{\psi_{j}}$. However, if the Q-function is in a nonlinear form, regression calibration can still considerably reduce the bias in the parameter estimation in Q-learning.  
	
	The following modified Q-learning algorithm with regression calibration details the estimation procedure: 
	
	1. Redefine the stage 2 history $\boldsymbol{H_{2}^{rc}} $ by replacing the surrogates with the regression calibration estimates.
	
	2. Parameterize the stage 2 Q-function  
	\begin{center}
		$Q_{2}(\boldsymbol{H_{2}^{rc}}, A_{2};\boldsymbol{\beta_{2}^{rc}}, \boldsymbol{\psi_{2}^{rc}})  = \boldsymbol{\beta_{2}^{rcT}H^{rc}_{20}} + \big(\boldsymbol{\psi_{2}^{rcT}H^{rc}_{21}}\big)A_{2}$.  
	\end{center}
	
	3. Apply OLS procedure and obtain the stage 2 estimator $(\boldsymbol{\hat{\beta}_{2}^{rc}},  \boldsymbol{\hat{\psi}_{2}^{rc}})$   
	\begin{center}
		$(\boldsymbol{\hat{\beta}_{2}^{rc}},  \boldsymbol{\hat{\psi}_{2}^{rc}})$ = $\underset{(\boldsymbol{\beta_{2}^{rc}}, \boldsymbol{\psi_{2}^{rc}})}{\arg\min}$ $\frac{1}{n}$ $\sum_{i=1}^{n}$ $\big(Y_{i} - Q_{2}(\boldsymbol{H_{i2}^{rc}}, A_{i2};\boldsymbol{\beta_{2}^{rc}}, \boldsymbol{\psi_{2}^{rc}})\big)^{2}$. 
	\end{center}
	
	4. Derive the stage 2 optimal treatment as $\hat{a}_{2}^{opt}$ = $\mathds{1}$$\big(\boldsymbol{\hat{\psi}^{rcT}_{2}h^{rc}_{21}} > 0\big)$.  
	
	5. Construct the pseudo-outcome for estimation at stage 1
	\begin{center}
		$\widetilde{Y}_{1}$ =  $\boldsymbol{\hat{\beta}^{rcT}_{2}}\boldsymbol{H^{rc}_{20}} +
		\big(\boldsymbol{\hat{\psi}^{rcT}_{2}}\boldsymbol{H^{rc}_{21}}\big)\mathds{1}\big(\boldsymbol{\hat{\psi}^{rcT}_{2}}\boldsymbol{H^{rc}_{21}} > 0\big)  $.    
	\end{center}
	
	6. Redefine the stage 1 history $\boldsymbol{H_{1}^{rc}} $ by replacing the surrogates with the regression calibration estimates.
	
	7. Parameterize the stage 1 Q-function
	\begin{center}
		$Q_{1}(\boldsymbol{H^{rc}_{1}}, A_{1};\boldsymbol{\beta^{rc}_{1}}, \boldsymbol{\psi^{rc}_{1}})  = \boldsymbol{\beta_{1}^{rcT}H^{rc}_{10}} + \big(\boldsymbol{\psi_{1}^{rcT}H^{rc}_{11}}\big)A_{1}$.  
	\end{center}

	8. Apply OLS procedure and obtain the stage 1 estimator $(\boldsymbol{\hat{\beta}_{1}^{rc}},  \boldsymbol{\hat{\psi}_{1}^{rc}})$
	\begin{center}
		$(\boldsymbol{\hat{\beta}_{1}^{rc}}, \boldsymbol{\hat{\psi}_{1}^{rc}})$ = $\underset{(\boldsymbol{\beta^{rc}_{1}}, \boldsymbol{\psi^{rc}_{1}})}{\arg\min}$ $\frac{1}{n}$ $\sum_{i=1}^{n}$ $\big(\widetilde{Y}_{i1} - Q_{1}(\boldsymbol{H^{rc}_{i1}}, A_{i1};\boldsymbol{\beta_{1}^{rc}}, \boldsymbol{\psi_{1}^{rc}})\big)^{2}$.  
	\end{center}  
	
	9. Derive the stage 1 optimal treatment as $\hat{a}_{1}^{opt}$ = $\mathds{1}\big(\boldsymbol{\hat{\psi}^{rcT}_{1}h^{rc}_{11}} > 0\big)$.

	\section{Simulation Studies} \label{sim}
	
	\subsection{One-Stage Estimation} \label{onechap1} 
	
	We begin with one-stage parameter estimation in Q-learning. Let $ X $ and $ Z $ be the error-prone and error-free covariates, respectively, generated from $ N $(1, 1). Instead of observing $ X $, two replicate surrogates $W_{1}$, $W_{2}$ are  generated as mismeasured version for $ X $, modeled by $W_{l}$ = $ X $ + $e_{l}$ ($l$ = 1, 2), where $e_{l} \sim N$(0, $\sigma^{2}$). $\overline{W}$ is the average value of $W_{1}$ and $W_{2}$. Treatment $A \in \{0, 1\}$ is generated from a Bernoulli distribution with probability $ P $($ A $ = 1) = 1/2. The outcome $ Y $ is generated by $ Y $ = 0.5 + $\beta_{z}Z$ + $\beta_{x}X$ + ($\psi_{10}$ + $\psi_{11}X$)$A$ + $\epsilon$, where ($\boldsymbol{\beta}$, $\boldsymbol{\psi}$) = $(\beta_{z}, \beta_{x}, \psi_{10}, \psi_{11})$ = (0.5, 1, 0.5, 1) and $\epsilon$ $\sim N$(0, 1), independent of each other and all the other variables.  
	
	Four estimators are considered and compared in each round of 500 simulations: (1) the true estimator $\boldsymbol{\hat{\psi}^{t}}$ obtained using the true covariate $X$, (2) the naive estimator $\boldsymbol{\hat{\psi}^{n}}$ obtained using a single surrogate $W_{1}$, (3) the naive estimator $\boldsymbol{\hat{\psi}^{nb}}$ obtained using the averaged surrogate $ \overline{W}$, (4) the regression calibration estimator $\boldsymbol{\hat{\psi}^{rc}}$ obtained using the regression calibration estimates $\hat{X}$. Analyses are conducted under two different sample sizes of $ n $ = 500 and $ n $ = 2000. In each setting, the bias, empirical standard error (SE), root mean square error (RMSE) and 95\% coverage rate (CR\%) of the estimators are calculated using the standard bootstrap with 200 bootstrap samples. The measurement error mechanism is assumed with $\sigma \in \{0.5, 0.7, 0.9\}$, which reflects a small, moderate and large measurement error on the true covariate $ X $. Numerical results are reported in Table $\ref{tab:one}$. The estimates of ($\boldsymbol{\beta}$, $\boldsymbol{\psi}$) for $ n $ = 2000 and $\sigma$ = 0.9 are visualized in Figure $\ref{fig:psi.2000.sigma3}$. 
	
	Table $\ref{tab:one}$ shows that ignoring the covariate measurement error leads to noticeably biased results, and the coverage rates are below the nominal level of 95\%. As the degree of measurement error increases, the biases are more severe. However, there is little impact on the estimation of $\beta_{z}$, which corresponds to the error-free covariate $  Z $. The regression calibration estimator presents a satisfactory performance in correcting for the effect with small biases and achieving coverage rates around 95\%. Its performance is also robust against the various magnitude of measurement error. Moreover, the sample size has effects on the performance of the methods as well. As the sample size becomes larger, the associated variability decreases in all the scenarios.     
	
	\subsection{Two-Stage Estimation with Linear Treatment-free Component} \label{two_lnchap1}
	
	This simulation study aims to investigate the effect of measurement error on the parameter estimation in DTR with two decision points. Let $X_{j}$ $\sim N$(1, 1) and $Z_{j} \sim N$(0.5, 1) be the error-prone and error-free covariates at stage $ j $ ($ j  $ = 1, 2), respectively. A treatment $A_{j} \in \{0, 1\}$ is assigned with probability $P(A_{j}$ = 1) = 1/2. In practice, the number of replicate surrogates may vary from person to person. To mimic this situation, we consider a scenario with 3 replicate surrogates $W_{j1}$, $W_{j2}$, $W_{j3}$, generated by $W_{jl} = X_{j}$ + $e _{jl}$ ($l$ = 1, 2, 3), where $e _{jl} \sim N$(0, $\sigma_{j}^{2}$). The degree of measurement error at stage $ j $ is reflected by $\sigma_{j}$, which is assumed to be known or estimated from a pilot study. Each patient is assumed to possess at least $W_{j1}$ and $W_{j2}$ as primary proxies while $W_{j3}$ may not be available. The degree of missingness in $W_{j3}$ is set to be 80\%. Let $ \overline{W}_{j} $ be an average value of $W_{j1}$, $W_{j2}$ and W$_{j3}$, $\overline{W}_{j}$ = $(W_{j1} + W_{j2} + W_{j3})/3$. The outcome is $ Y $ = $X_{1}$ + $Z_{1}$ + $X_{2}$ + $Z_{2}$ + ($\psi_{10}$ + $\psi_{11}X_{1}$)$A_{1}$ + ($\psi_{20}$ + $\psi_{21}X_{2}$)$A_{2}$  + $\epsilon$, where $\boldsymbol{\psi}$ = ($\psi_{20}$, $\psi_{21}$, $\psi_{10}$, $\psi_{11}$) = (0.5, -1, 0.5, -1) and $\epsilon \sim N$(0, 1) is generated independently of each other and of all other variables. In each round of 500 simulations, a dataset with the size of 2000 patients is generated.   
	
	Four estimators are compared in each stage to evaluate the performance of the  regression calibration method: (1) the true estimator $\boldsymbol{\hat{\psi}^{t}}$ obtained using the true covariate $X_{j}$, (2) the naive estimator $\boldsymbol{\hat{\psi}^{n}}$ obtained using a single surrogate $W_{j1}$, (3) the naive estimator $\boldsymbol{\hat{\psi}^{nb}}$ obtained using the averaged surrogate $\overline{W}_{j}$, (4) the regression calibration estimator $\boldsymbol{\hat{\psi}^{rc}}$ obtained using the regression calibration estimates $\hat{X}_{j}$. The degree of measurement error $\sigma_{j}$ is specified as 0.5, 0.7 and 0.9. Results for the bias, SE, RMSE and CR\% of $\boldsymbol{\hat{\psi}}$ computed using the standard bootstrap with 200 bootstrap samples are reported in Table $\ref{tab:two_linear}$. Figure $\ref{fig:two9}$ provides the visualized parameter estimates under $\sigma_{j}$ = 0.9. 
	
	Similar to the findings in one-stage estimation, both the naive blip estimates $\boldsymbol{\hat{\psi}^{n}}$ and $\boldsymbol{\hat{\psi}^{nb}}$ are biased due to the ignorance of the covariate measurement error. The biases of the naive estimators exacerbate as the degree of measurement error increases. On the contrary, the regression calibration estimator $\boldsymbol{\hat{\psi}^{rc}}$ yields small biases, and the coverage rates are close to the nominal level of 95\%. The performance of the regression calibration estimator is also shown to be robust against the magnitude of measurement error across the two stages.

	\subsection{Two-Stage Estimation with Nonlinear Treatment-free Component} \label{two_nonlchap1} 
	
	In this section, we explore the measurement error effect on the estimation of blip parameters and optimal DTR in a nonlinear outcome model with two decision points. The data generation mechanism is the same as the one in (\ref{two_lnchap1}) except that the outcome model is given by $ Y $ = $ f $($X_{1}$) + $Z_{1}$ + $ f $($X_{2}$) + $Z_{2}$ + ($\psi_{10} + \psi_{11}X_{1}$)$A_{1}$ + $(\psi_{20} + \psi_{21}X_{2}$)$A_{2}$  + $\epsilon$, where $\boldsymbol{\psi}$ = ($\psi_{20}$, $\psi_{21}$, $\psi_{10}$, $\psi_{11}$) = (0.5, -1, 0.5, -1), and $\epsilon \sim N$(0, 1) is generated and independent of everything else. In the outcome model, three nonlinear functions are considered for $X_{j}$: (1) $f$($X_{j}$) = $X_{j}$ + $X_{j}^{3}$ (cubic), (2) $ f $($X_{j}$) = $X_{j}$ + $e^{X_{j}}$ (exponential), (3) $ f $($X_{j}$) = $X_{j}$ + sin($X_{j}^{2}$) + cos($X_{j}^{2}$) (complex). 
	
	We continue the analysis with four estimators $\boldsymbol{\hat{\psi}^{t}}$, $\boldsymbol{\hat{\psi}^{n}}$, $\boldsymbol{\hat{\psi}^{nb}}$ and $\boldsymbol{\hat{\psi}^{rc}}$. The measurement error $\sigma_{j}$ is chosen from a range of \{0.5, 0.7, 0.9\}. Table $\ref{tab:two_nonlinear}$ displays the results for the bias, SE, RMSE and CR\% of $\boldsymbol{\hat{\psi}}$ over various measurement errors in each nonlinear case. The blip estimates for three nonlinear examples under ($\sigma_{2}$, $\sigma_{1}$) = (0.9, 0.9) are visualized in Figures $\ref{fig:two_cubic3}$, $\ref{fig:two_exp3}$ and $\ref{fig:two_comp3}$. We repeat the set of simulations 500 times.   
	
	In general, the results in Table $\ref{tab:two_nonlinear}$ reveal a larger measurement error effect with larger biases and standard errors in the nonlinear case than in the linear case. In comparison, the regression calibration method remains effective and robust, though it produces slightly less reduced biases in this setting compared with those in (\ref{two_lnchap1}), especially for the estimation of $\boldsymbol{\psi_{1}}$. Moreover, in the three scenarios, the model containing the complex function in the treatment-free component is more sensitive to measurement error.

	\subsection{Predicted Optimal DTR} \label{pred_chap1} 
	
	In this section, we explore the effect of measurement error on the predicted optimal treatment decision rules by evaluating the proportion of optimally treated patients across two stages. Although there is an argument about the necessity of modeling measurement error in a predictive setting \citep{carroll2006measurement}, considering the importance of correctly identifying and recommending the optimal treatments to the patients, it's worth looking into the role of measurement error in predicting the optimal DTR in Q-learning.   
	
	The analysis follows the simulation design (\ref{two_lnchap1}) and is done with the training data of 2000 patients and test data of 5000 patients. We first use the training data to produce three estimators ($\boldsymbol{\hat{\psi}^{n}}$, $\boldsymbol{\hat{\psi}^{nb}}$, $\boldsymbol{\hat{\psi}^{rc}}$) with a single surrogate $W_{j1}$, averaged surrogate $\overline{W}_{j}$ and regression calibration estimates $\hat{X}_{j}$, respectively. Then, we use the test data to find the prediction accuracy of optimal DTR, which is measured by the proportion of the patients whose optimal treatments are correctly identified in the test data at stage 2 and/or stage 1. 
	
	In each stage, six scenarios are considered to predict the optimal DTR using (1) the naive estimator $\boldsymbol{\hat{\psi}^{n}}$ and true covariate $X_{j}$ (nt), (2) the naive estimator $\boldsymbol{\hat{\psi}^{nb}}$ and true covariate $X_{j}$ (nbt), (3) the regression calibration estimator $\boldsymbol{\hat{\psi}^{rc}}$ and true covariate $X_{j}$ (ct), (4) the naive estimator $\boldsymbol{\hat{\psi}^{n}}$ and a single surrogate $W_{j1}$ (nn), (5) the naive estimator $\boldsymbol{\hat{\psi}^{nb}}$ and averaged surrogate $\overline{W}_{j}$ (nbnb), (6) the regression calibration estimator $\boldsymbol{\hat{\psi}^{rc}}$ and regression calibration estimates $\hat{X}_{j}$ (cc), respectively. The first three scenarios aim to examine the measurement error effect on the prediction accuracy using the true covariates in the test data, while the last three evaluate the measurement error effect using the surrogates and corrected estimates in the test data. A total of 500 runs are simulated for each scenario. Numerical results are summarized in Table \ref{tab:predoptDTR}. 
	
	Table \ref{tab:predoptDTR} shows that the existence of measurement error leads to a remarkable degradation of the prediction accuracy of optimal DTR, and it achieves the lowest prediction accuracy when a single surrogate is used. However, the regression calibration method outperforms the naive method and significantly improves the prediction accuracy in all the scenarios. In the last two scenarios (nbnb and cc), the regression calibration method yields similar accuracy results to the naive method, indicating that the worst scenario of using the regression calibration method is comparable to that of using the naive method.

	\subsection{Predicted Optimal Value Function}  \label{valuefun}
	
	In this study, we evaluate the measurement error effect on the predicted optimal value function, which is the expected outcome under the optimal treatment regimes. The data generation mechanism follows (\ref{pred_chap1}), and we continue with the three estimators ($\boldsymbol{\hat{\psi}^{n}}$, $\boldsymbol{\hat{\psi}^{nb}}$, $\boldsymbol{\hat{\psi}^{rc}}$) obtained from the training data. We use the test data to predict the value functions under (1) the true optimal DTR (opt), (2) the optimal DTR estimated using $\boldsymbol{\hat{\psi}^{n}}$ and $X_{j}$ (nt), (3) the optimal DTR estimated using $\boldsymbol{\hat{\psi}^{nb}}$ and $X_{j}$ (nbt), (4) the optimal DTR estimated using $\boldsymbol{\hat{\psi}^{rc}}$ and $X_{j}$ (ct),  (5) the optimal DTR estimated using $\boldsymbol{\hat{\psi}^{n}}$ and $W_{j1}$ (nn),  (6) the optimal DTR estimated using $\boldsymbol{\hat{\psi}^{nb}}$ and $\overline{W}_{j}$ (nbnb),  (7) the optimal DTR estimated using $\boldsymbol{\hat{\psi}^{rc}}$ and $\hat{X}_{j}$ (cc). Simulations are repeated 500 times. For each scenario, the average value function is computed and reported in Table \ref{tab:pred.vfun}, along with its standard deviations.
	
	In Table \ref{tab:pred.vfun}, we see that the measurement error effect is pronounced in terms of value function estimation under the optimal DTR. By comparison, the naive method generally yields lower value function estimates, and the optimal value function achieves the lowest value with a single surrogate being used, as expected. The regression calibration method, however, improves the estimated optimal value function, even comparable to the true optimal value function when the true covariate is used.

	\section{Application to STAR*D Data}  \label{da}
	
	To illustrate the proposed correction method, we analyze the data arising from the Sequenced Treatment Alternatives to Relieve Depression study \citep{rush200316, rush2004sequenced}. The STAR*D study was designed as a multisite, multistage randomized controlled trial. It aimed to evaluate the effect of treatments for patients who suffered from major depressive disorder. The severity of depressive disorder was measured by the Quick Inventory of Depressive Symptomatology score, which was assessed by both patients (QIDS-S) and clinicians (QIDS-C). The entire study possessed four levels, in which one or a combination of treatments was assigned to the patients. At level 1, all of the patients were prescribed citalopram (CIT). At the end of level 1, if patients had QIDS $\leq$ 5, they achieved remission and were removed from the study but those who otherwise entered level 2. They were again randomized into one of the seven treatment options: either switching from CIT to one of four other treatment options (venlafaxine[VEN], sertraline[SER], bupropion[BUP], and cognitive therapy[CT]) or augmenting CIT with one of three treatments (BUP, CT and buspirone[BUS]). Then, at the end of level 2, they were again assessed with the QIDS score, and those who failed to achieve remission (QIDS $\leq$ 5) entered level 3. In level 3, they were randomized to receive either one of two new treatments (lithium[Li] or thyroid hormone[THY]) or one of two augmented treatment options (mirtazapine[MIRT], nortriptyline[NTP]). The QIDS score for remission was evaluated at the end of level 3. 
	
	In the literature, depression is found to be significantly associated with functional impairment \citep{greer2010defining}. Patients with major depressive disorder were shown to have considerable deficits in the physical and social functioning \citep{lin2014depression, trivedi2013increase}. The importance of developing individualized treatments for patients with a major depressive disorder was pointed out to improve their long-term functioning \citep{ishak2016patient}. The perceived functional impairment is measured at each level of the STAR*D study by the Work and Social Adjustment Scale (WSAS) score, which reflects the functioning aspects of the work, home management, social activities, private activities, and relationships with others.    
	
	We follow the criteria in the literature \citep{chakraborty2013statistical, chakraborty2013inference, wallace2019model} to select the data, where the two-stage DTR is considered by combining level 2 and level 2A as the first stage and treating level 3 as the second stage. The stage $ j $ treatment $A_{j}$ is coded based on whether the treatment involves selective serotonin reuptake inhibitor ($A_{j}$ = 1) or not ($A_{j}$ = 0). Three tailoring variables are considered, $Q_{j}$: the QIDS-C score measured at the beginning of each level $ j $, $S_{j}$: the QIDS slope, the change in QIDS-C divided by the time in the previous level, and $P_{j}$: the patients' preference indicating whether they wished to switch previous treatment ($P_{j}$ = 1), to augment previous treatment or have no preference ($P_{j}$ = 0). The outcome of interest is defined as the negative WSAS score across two stages
	\begin{center}
		$ Y = R_{1} \cdot Y_{1} + (1 - R_{1}) \cdot \frac{1}{2}(Y_{1} + Y_{2}), $
	\end{center}
	where $Y_{1} $ and $Y_{2} $ are the negative WSAS scores observed at the end of stage 1 and stage 2, and $R_{1}$ is an indicator of whether the patients achieved remission ($R_{1}$ = 1) or not ($R_{1}$ = 0) at the end of stage 1. The selected data contain 1438 patients at stage 1, of whom 377 patients have entered the stage 2.    
	
	The previous analyses of the STAR*D data often assume that the QIDS-C score is error-free, which is usually not the case in practice. The measurement error effect on sequential optimal treatment rules was studied, assuming that the true QIDS score was unknown and both the QIDS-C score and QIDS-S score were considered as the repeated measurements of the true underlying QIDS score \citep{spicker2020measurement}. In this work, we are interested in estimating the optimal treatment decision rules using Q-learning that maximize the negative WASA score, provided that the QIDS score is subject to measurement error. We compare three estimators, including two naive estimators using the QIDS-C score or QIDS-S score as the tailoring variable and the regression calibration estimator using the corrected estimates computed by treating QIDS-C and QIDS-S as repeated measurements of QIDS. The analysis results of the parameter estimates, bootstrap standard error with 200 bootstrap samples, and 95\% confidence interval are summarized in Table \ref{tab:stard}. 
	
	In Table \ref{tab:stard}, the parameter estimates of each stage vary remarkably between the naive method and the regression calibration method, leading to different optimal treatment decision rules. More importantly, the results show that the significance of the tailoring variable differs between these two methods. The patients' preference to switch treatment and the QIDS score have significant treatment effects in the interaction with the second stage treatment when the QIDS-S score is used. However, by using the regression calibration estimates, no significant term is observed across the two stages. It emphasizes that the measurement error effect is not negligible in an error-prone setting since it is possible to alter the estimation of optimal treatment decision rules and the significance of the tailoring variable.

	\section{Discussion} \label{conclude}
	
	In this paper, we study the Q-learning with covariate measurement error, provided that repeated measurements of the true covariate are available. Q-learning is one of the most popular regression-based methods to estimate an optimal DTR, but it has not been studied in an error-prone setting. We fill the gap by exploring the impact of covariate measurement error on the performance of Q-learning and demonstrate the severe consequence of ignoring the measurement error in the estimation. Therefore, we propose the regression calibration method to correct the measurement error in Q-learning. By making use of the available replicate surrogates, the regression calibration method yields consistent estimates of blip parameters for the linear outcome model and still provides substantial bias reduction for the nonlinear outcome model.  
	
	Simulation studies are conducted to evaluate the performance of the proposed method from different aspects. The results demonstrate that the estimation of blip parameter is biased using the naive method in single-stage and multi-stage settings. The simulation showcases the competency of the regression calibration method by considerably reducing the bias, especially in linear Q-learning. The performance of the regression calibration method is stable and robust against various magnitudes of measurement error. For nonlinear outcome models, the regression calibration still provides substantial bias reduction and achieves favorable results.     
	
	Another important topic discussed in this article is to assess the performance of the proposed method from a predictive perspective. We predict the future optimal treatment decision rules in various settings using both the naive method and the proposed method. It turns out that using a single mismeasured covariate leads to the worst prediction accuracy of optimal DTR among all methods, while the regression calibration method improves the prediction accuracy even when the degree of measurement error is large. Moreover, we also compare the naive method and regression calibration method in terms of value function estimation. The optimal value function and its variability estimated from the naive method are generally lower, while the regression calibration method enhances the optimal value function comparable to the true optimal value function.   
	
	For illustration, we apply the proposed method to the STAR*D data . The self-reported and clinician-evaluated QIDS scores in the STAR*D data are error-prone, which is rarely considered in the literature. Our results exhibit the importance of taking the measurement error into account in estimating the sequential optimal treatment decision rules, which can be altered otherwise. 
	
	The study of measurement error in dynamic treatment regimes is a new and challenging topic. As the performance of the regression calibration is mostly favorable in linear models, it is of interest to consider other correction methods for highly nonlinear models in Q-learning. Moreover, the DTR method studied in this paper is based on a continuous outcome, but it remains unexplored for other classes of outcomes, which is worth investigating in the future.

	\paragraph{Data Availability Statement} 
	The STAR*D study data are available with restriction from the National Institute of Mental Health Data Archive (\url{https://nda.nih.gov/edit_collection.html?id=2148}).

	\bibliographystyle{apalike}
	\bibliography{QlearnMe_arXiv}
	
		
	
	\begin{table} 
		\caption{One-stage estimates of blip parameters $(\psi_{10}, \psi_{11})$}
		\label{tab:one}
		\centering
		\begin{tabular}{ccrrccrrrcrr}
			\hline
			{} & {} & {} & \multicolumn{4}{@{}c@{}}{$\psi_{10}$}  & \multicolumn{4}{@{}c@{}}{$\psi_{11}$}  \\ 
			\cmidrule{4-7} \cmidrule{8-11} 
			n & $\sigma$ & $ \boldsymbol{\hat{\psi}} $ & \multicolumn{1}{c}{Bias} & \multicolumn{1}{c}{SE} & \multicolumn{1}{c}{RMSE} & \multicolumn{1}{c}{CR\%} & \multicolumn{1}{c}{Bias} & \multicolumn{1}{c}{SE} & \multicolumn{1}{c}{RMSE} & \multicolumn{1}{c}{CR\%}     \\
			\hline
			500 & {} & $\boldsymbol{\hat{\psi}^{t}}$ & 0.002	& 0.133	& 0.133	& 96.8	& -0.002	& 0.095	& 0.095	& 95.2 \\
			{} & 0.5 & $\boldsymbol{\hat{\psi}^{n}} $ &  0.210	& 0.150	& 0.258	& 72.0	& -0.209	& 0.101	& 0.232	& 44.6  \\
			{} & {} & $\boldsymbol{\hat{\psi}^{nb}}$ &  0.114	& 0.143	& 0.183	& 88.6	& -0.116	& 0.099	& 0.152	& 76.4  \\
			{} & {} & $\boldsymbol{\hat{\psi}^{rc}} $ &  0.004	& 0.153	& 0.153	& 94.8	& -0.005	& 0.111	& 0.111	& 95.6  \\
			{} & 0.7 & $\boldsymbol{\hat{\psi}^{n}}$ & 0.338	& 0.157	& 0.373	& 42.4	& -0.337	& 0.100	& 0.351	& 9.0  \\
			{} & {} & $\boldsymbol{\hat{\psi}^{nb}}$ &  0.204	& 0.149	& 0.253	& 73.6	& -0.203	& 0.100	& 0.227	& 48.8  \\
			{} & {} & $\boldsymbol{\hat{\psi}^{rc}} $ & 0.009	& 0.167	& 0.167	& 93.8	& -0.009	& 0.124	& 0.124	& 95.2  \\
			{} & 0.9 & $\boldsymbol{\hat{\psi}^{n}}$ & 0.463	& 0.163	& 0.491	& 20.4	& -0.460	& 0.097	& 0.470	& 0.6  \\
			{} & {} & $\boldsymbol{\hat{\psi}^{nb}}$ &  0.299	& 0.155	& 0.337	& 49.4	& -0.300	& 0.100	& 0.316	& 17.2  \\
			{} & {} & $\boldsymbol{\hat{\psi}^{rc}} $ & 0.012	& 0.185	& 0.185	& 95.2	& -0.013	& 0.141	& 0.141	& 94.2  \\
			\midrule
			2000 & {} & $\boldsymbol{\hat{\psi}^{t}}$ &  0.001	& 0.066	& 0.066	& 96.0	& 0.000	& 0.047	& 0.047	& 96.2  \\
			{} & 0.5 & $ \boldsymbol{\hat{\psi}^{n}} $ &  0.194	& 0.075	& 0.208	& 24.0	& -0.195	& 0.050	& 0.201	& 2.8  \\
			{} & {} & $\boldsymbol{\hat{\psi}^{nb}} $ & 0.107	& 0.072	& 0.129	& 68.0	& -0.107	& 0.049	& 0.117	& 39.6  \\
			{} & {} & $\boldsymbol{\hat{\psi}^{rc}} $ &  -0.005	& 0.076	& 0.076	& 94.6	& 0.005	& 0.055	& 0.055	& 95.4  \\
			{} & 0.7 & $ \boldsymbol{\hat{\psi}^{n}} $ & 0.330	& 0.079	& 0.339	& 1.8	& -0.328	& 0.050	& 0.332	& 0.0  \\
			{} & {} & $\boldsymbol{\hat{\psi}^{nb}} $ &  0.200	& 0.075	& 0.213	& 24.4	& -0.197	& 0.050	& 0.203	& 2.4 \\
			{} & {} & $\boldsymbol{\hat{\psi}^{rc}} $ &  0.003	& 0.084	& 0.084	& 94.6	& 0.000	& 0.062	& 0.062	& 94.4  \\
			{} & 0.9 & $ \boldsymbol{\hat{\psi}^{n}} $ &  0.445	& 0.081	& 0.452	& 0.0	& -0.446	& 0.049	& 0.449	& 0.0 \\
			{} & {} & $\boldsymbol{\hat{\psi}^{nb}} $ & 0.286	& 0.077	& 0.296	& 4.4	& -0.287	& 0.050	& 0.291	& 0.0 \\
			{} & {} & $\boldsymbol{\hat{\psi}^{rc}} $ &  -0.003	& 0.092	& 0.092	& 93.8	& 0.002	& 0.070	& 0.070	& 92.2 \\
			\hline
		\end{tabular} 
	\end{table}

	\begin{sidewaystable}[hbt!]
		\setlength\tabcolsep{2pt} 
		\caption{Two-stage estimates of blip parameters ($\psi_{20}$, $\psi_{21}$, $\psi_{10}$, $\psi_{11}$) with linear treatment-free component}
		\label{tab:two_linear}
		\centering
		\begin{adjustbox}{width=0.8\textwidth}
			\begin{tabular}{rrrrccrrccrrccrrccr}
			\midrule
			{} & {} & {}  &  \multicolumn{4}{c}{$\psi_{20}$} & \multicolumn{4}{c}{$\psi_{21}$}  &  \multicolumn{4}{c}{$\psi_{10}$} & \multicolumn{4}{c}{$\psi_{11}$}   \\
			\cmidrule{4-7} \cmidrule{8-11} \cmidrule{12-15} \cmidrule{16-19} 
			$\sigma_{2}$ & $\sigma_{1}$ & $ \boldsymbol{\hat{\psi}} $ & \multicolumn{1}{c}{Bias} & \multicolumn{1}{c}{SE} & \multicolumn{1}{c}{RMSE} & \multicolumn{1}{c}{CR\%} & \multicolumn{1}{c}{Bias} & \multicolumn{1}{c}{SE} & \multicolumn{1}{c}{RMSE} & \multicolumn{1}{c}{CR\%} & \multicolumn{1}{c}{Bias} & \multicolumn{1}{c}{SE} & \multicolumn{1}{c}{RMSE} & \multicolumn{1}{c}{CR\%} & \multicolumn{1}{c}{Bias} & \multicolumn{1}{c}{SE} & \multicolumn{1}{c}{RMSE} & \multicolumn{1}{c}{CR\%}    \\
			\midrule 
			{} & {} & $ \boldsymbol{\hat{\psi}^{t}} $  & 0.001	& 0.072	& 0.072	& 97.6	& -0.002	& 0.052	& 0.052	& 97.8	& 0.001	& 0.105	& 0.105	& 94.6	& 0.001	& 0.075	& 0.075	& 95.2  \\
			0.5  & 0.5 & $  \boldsymbol{\hat{\psi}^{n}}  $   & -0.196	& 0.072	& 0.209	& 20.2	& 0.197	& 0.048	& 0.203	& 1.0	& -0.204	& 0.101	& 0.228	& 46.4	& 0.201	& 0.067	& 0.212	& 12.8  \\
			{} & {} & $\boldsymbol{\hat{\psi}^{nb}} $ & -0.100	& 0.072	& 0.123	& 73.4	& 0.100	& 0.050	& 0.112	& 49.4	& -0.109	& 0.103	& 0.150	& 81.2	& 0.106	& 0.071	& 0.127	& 67.4   \\
			{} & {} & $ \boldsymbol{\hat{\psi}^{rc}} $ & 0.006	& 0.076	& 0.076	& 96.4	& -0.006	& 0.056	& 0.056	& 96.0	& -0.004	& 0.109	& 0.109	& 95.8	& 0.001	& 0.079	& 0.079	& 97.0  \\
			{} & 0.7& $  \boldsymbol{\hat{\psi}^{n}}  $   & -0.201	& 0.073	& 0.214	& 20.6	& 0.201	& 0.049	& 0.207	& 1.0	& -0.335	& 0.098	& 0.349	& 7.2	& 0.330	& 0.062	& 0.336	& 0.0  \\
			{} & {} & $ \boldsymbol{\hat{\psi}^{nb}}  $ & -0.104	& 0.073	& 0.127	& 70.4	& 0.104	& 0.051	& 0.116	& 46.6	& -0.193	& 0.101	& 0.217	& 52.2	& 0.189	& 0.068	& 0.201	& 19.6  \\
			{} & {} & $ \boldsymbol{\hat{\psi}^{rc}}  $ & 0.001	& 0.077	& 0.077	& 96.0	& -0.001	& 0.056	& 0.056	& 97.2	& -0.007	& 0.112	& 0.113	& 95.6	& 0.004	& 0.084	& 0.084	& 96.6  \\
			{} & 0.9 & $  \boldsymbol{\hat{\psi}^{n}} $   & -0.200	& 0.074	& 0.213	& 20.4	& 0.200	& 0.050	& 0.206	& 1.2	& -0.448	& 0.095	& 0.457	& 0.2	& 0.449	& 0.057	& 0.452	& 0.0  \\
			{} & {} & $ \boldsymbol{\hat{\psi}^{nb}}  $ & -0.105	& 0.073	& 0.128	& 72.2	& 0.104	& 0.051	& 0.116	& 45.6	& -0.275	& 0.099	& 0.292	& 21.0	& 0.277	& 0.065	& 0.284	& 1.6  \\
			{} & {} & $ \boldsymbol{\hat{\psi}^{rc}}  $ & 0.000	& 0.078	& 0.078	& 97.4	& 0.000	& 0.057	& 0.057	& 97.4	& -0.001	& 0.116	& 0.116	& 94.4	& 0.003	& 0.089	& 0.089	& 94.8 \\
			0.7  & 0.5 & $  \boldsymbol{\hat{\psi}^{n}}  $   & -0.332	& 0.070	& 0.340	& 0.0	& 0.330	& 0.045	& 0.333	& 0.0	& -0.199	& 0.101	& 0.223	& 49.4	& 0.199	& 0.068	& 0.211	& 16.4  \\
			{} & {} & $ \boldsymbol{\hat{\psi}^{nb}}  $ & -0.188	& 0.071	& 0.201	& 23.8	& 0.186	& 0.048	& 0.192	& 1.4	& -0.103	& 0.103	& 0.146	& 83.0	& 0.103	& 0.072	& 0.126	& 70.8  \\
			{} & {} & $ \boldsymbol{\hat{\psi}^{rc}} $  & -0.002	& 0.079	& 0.079	& 95.6	& 0.000	& 0.059	& 0.059	& 96.0	& 0.002	& 0.109	& 0.109	& 95.8	& -0.001	& 0.080	& 0.080	& 95.8 \\
			{} & 0.7 & $  \boldsymbol{\hat{\psi}^{n}}  $   & -0.334	& 0.071	& 0.342	& 0.2	& 0.331	& 0.045	& 0.334	& 0.0	& -0.326	& 0.098	& 0.340	& 8.0	& 0.327	& 0.062	& 0.333	& 0.0   \\
			{} & {} & $ \boldsymbol{\hat{\psi}^{nb}} $ &  -0.190	& 0.072	& 0.203	& 24.0	& 0.187	& 0.048	& 0.193	& 1.8	& -0.186	& 0.101	& 0.212	& 53.6	& 0.186	& 0.068	& 0.198	& 22.2 \\
			{} & {} & $ \boldsymbol{\hat{\psi}^{rc}} $  & -0.004	& 0.080	& 0.080	& 96.0	& 0.000	& 0.059	& 0.059	& 97.8	& 0.000	& 0.112	& 0.112	& 95.8	& 0.000	& 0.084	& 0.084	& 95.6  \\
			{} & 0.9 & $  \boldsymbol{\hat{\psi}^{n}}  $   & -0.328	& 0.072	& 0.336	& 0.4	& 0.329	& 0.046	& 0.332	& 0.0	& -0.449	& 0.095	& 0.459	& 0.0	& 0.446	& 0.057	& 0.450	& 0.0  \\
			{} & {} & $ \boldsymbol{\hat{\psi}^{nb}} $ & -0.185	& 0.073	& 0.199	& 24.8	& 0.185	& 0.049	& 0.192	& 2.4	& -0.277	& 0.100	& 0.295	& 22.0	& 0.275	& 0.065	& 0.283	& 0.8  \\
			{} & {} & $ \boldsymbol{\hat{\psi}^{rc}}  $ & 0.001	& 0.081	& 0.081	& 97.0	& -0.001	& 0.061	& 0.061	& 97.0	& -0.002	& 0.117	& 0.117	& 94.0	& 0.001	& 0.089	& 0.089	& 96.0  \\
			0.9  & 0.5 & $  \boldsymbol{\hat{\psi}^{n}}  $  & -0.449	& 0.068	& 0.454	& 0.0	& 0.447	& 0.041	& 0.449	& 0.0	& -0.203	& 0.101	& 0.227	& 48.2	& 0.201	& 0.068	& 0.212	& 16.4 \\
			{} & {} & $\boldsymbol{\hat{\psi}^{nb}} $  & -0.273	& 0.070	& 0.282	& 2.0	& 0.272 & 	0.046	& 0.276	& 0.0	& -0.106	& 0.103	& 0.148	& 82.8	& 0.104	& 0.071	& 0.126	& 69.2 \\
			{} & {} & $ \boldsymbol{\hat{\psi}^{rc}} $  & 0.004	& 0.082	& 0.082	& 97.0	& -0.005	& 0.063	& 0.063	& 95.6	& -0.002	& 0.109	& 0.109	& 95.0	& 0.000	& 0.080	& 0.080	& 96.0  \\
			{} & 0.7 & $ \boldsymbol{\hat{\psi}^{n}}  $   &  -0.451	& 0.069	& 0.456	& 0.0	& 0.450	& 0.042	& 0.452	& 0.0	& -0.330	& 0.098	& 0.344	& 7.2	& 0.330	& 0.062	& 0.336	& 0.0 \\
			{} & {} & $ \boldsymbol{\hat{\psi}^{nb}} $ & -0.276	& 0.070	& 0.285	& 2.0	& 0.276	& 0.046	& 0.280	& 0.0	& -0.186	& 0.101	& 0.212	& 55.0	& 0.187	& 0.068	& 0.199	& 21.2  \\
			{} & {} & $ \boldsymbol{\hat{\psi}^{rc}} $ & -0.001	& 0.083	& 0.083	& 97.2	& 0.001	& 0.064	& 0.064	& 95.4	& 0.000	& 0.113	& 0.113	& 95.0	& 0.001	& 0.084	& 0.084	& 97.0 \\
			{} & 0.9 & $  \boldsymbol{\hat{\psi}^{n}}  $   & -0.449	& 0.070	& 0.454	& 0.0	& 0.448	& 0.042	& 0.450	& 0.0	& -0.446	& 0.095	& 0.456	& 0.4	& 0.445	& 0.057	& 0.449	& 0.0 \\
			{} & {} & $ \boldsymbol{\hat{\psi}^{nb}}  $ & -0.274	& 0.071	& 0.283	& 2.6	& 0.274	& 0.047	& 0.278	& 0.0	& -0.273	& 0.100	& 0.291	& 21.8	& 0.272	& 0.065	& 0.280	& 1.2 \\
			{} & {} & $ \boldsymbol{\hat{\psi}^{rc}}  $ & 0.000	& 0.084	& 0.084	& 96.6	& 0.000	& 0.064	& 0.064	& 97.2	& 0.002	& 0.117	& 0.117	& 94.8	& -0.003	& 0.089	& 0.089	& 95.6 \\
			\midrule
		\end{tabular}
			\end{adjustbox}
\end{sidewaystable}

	\begin{sidewaystable}[hbt!]
		\setlength\tabcolsep{2pt} 
		\caption{Two-stage estimates of blip parameters ($\psi_{20}$, $\psi_{21}$, $\psi_{10}$, $\psi_{11}$) with nonlinear treatment-free component}
		\label{tab:two_nonlinear}
		\centering
		\begin{adjustbox}{width=0.8\textwidth}
			\begin{tabular}{ccrrccrrccrrccrrccr}
				\midrule
			{} & {} & {}  &  \multicolumn{4}{c}{$\psi_{20}$} & \multicolumn{4}{c}{$\psi_{21}$}  &  \multicolumn{4}{c}{$\psi_{10}$} & \multicolumn{4}{c}{$\psi_{11}$}   \\
			\cmidrule{4-7} \cmidrule{8-11} \cmidrule{12-15} \cmidrule{16-19} 
			Scenario & ($\sigma_{2}$, $\sigma_{1}$) & $ \boldsymbol{\hat{\psi}} $ & \multicolumn{1}{c}{Bias} & \multicolumn{1}{c}{SE} & \multicolumn{1}{c}{RMSE} & \multicolumn{1}{c}{CR\%} & \multicolumn{1}{c}{Bias} & \multicolumn{1}{c}{SE} & \multicolumn{1}{c}{RMSE} & \multicolumn{1}{c}{CR\%} & \multicolumn{1}{c}{Bias} & \multicolumn{1}{c}{SE} & \multicolumn{1}{c}{RMSE} & \multicolumn{1}{c}{CR\%} & \multicolumn{1}{c}{Bias} & \multicolumn{1}{c}{SE} & \multicolumn{1}{c}{RMSE} & \multicolumn{1}{c}{CR\%}   \\
				\midrule 
				cub & {} & $\boldsymbol{\hat{\psi}^{t}}$  & 0.002	& 0.068	& 0.068	& 96.4	& -0.001	& 0.049	& 0.049	& 98.0	& 0.017	& 0.540	& 0.540	& 97.4	& -0.013	& 0.381	& 0.382	& 95.6   \\
				{} & (0.5, 0.5) & $  \boldsymbol{\hat{\psi}^{n}}  $ &  -0.184	& 0.315	& 0.365	& 91.8	& 0.180	& 0.332	& 0.378	& 91.6	& -0.215	& 0.539	& 0.580	& 94.4	& 0.182	& 0.443	& 0.479	& 92.6  \\
				{} & {} & $ \boldsymbol{\hat{\psi}^{nb}}  $  &  -0.085	& 0.246	& 0.260	& 94.0	& 0.082	& 0.268	& 0.280	& 94.4	& -0.118	& 0.538	& 0.551	& 95.2	& 0.094	& 0.425	& 0.436	& 92.0   \\
				{} & {} & $ \boldsymbol{\hat{\psi}^{rc}} $  &  0.023	& 0.262	& 0.263	& 94.8	& -0.025	& 0.295	& 0.296	& 95.6	& -0.010	& 0.574	& 0.574	& 95.2	& -0.014	& 0.475	& 0.475	& 93.2  \\
				{} & (0.7, 0.7) & $  \boldsymbol{\hat{\psi}^{n}}  $ &  -0.337	& 0.367	& 0.498	& 82.8	& 0.316	& 0.368	& 0.485	& 85.0	& -0.358	& 0.528	& 0.638	& 89.8	& 0.331	& 0.440	& 0.551	& 88.6    \\
				{} & {} & $ \boldsymbol{\hat{\psi}^{nb}}  $  &  -0.179	& 0.306	& 0.354	& 91.4	& 0.172	& 0.325	& 0.368	& 89.6	& -0.211	& 0.539	& 0.579	& 93.6	& 0.184	& 0.443	& 0.480	& 91.2   \\
				{} & {} & $  \boldsymbol{\hat{\psi}^{rc}}  $  & 0.012	& 0.349	& 0.350	& 96.0	& -0.019	& 0.395	& 0.395	& 93.8	& -0.022	& 0.606	& 0.606	& 95.2	& -0.004	& 0.539	& 0.539	& 93.8   \\
				{} & (0.9, 0.9) & $  \boldsymbol{\hat{\psi}^{n}} $ &  -0.453	& 0.401	& 0.605	& 78.8	& 0.475	& 0.370	& 0.602	& 75.0	& -0.455	& 0.517	& 0.689	& 86.8	& 0.455	& 0.416	& 0.616	& 77.8    \\
				{} & {} & $ \boldsymbol{\hat{\psi}^{nb}}  $  & -0.271	& 0.352	& 0.444	& 89.8	& 0.283	& 0.362	& 0.460	& 87.8	& -0.274	& 0.536	& 0.602	& 92.2	& 0.286	& 0.448	& 0.531	& 87.4  \\
				{} & {} & $ \boldsymbol{\hat{\psi}^{rc}} $  &  -0.002	& 0.435	& 0.435	& 96.4	& 0.014	& 0.489	& 0.490	& 95.4	& 0.002	& 0.645	& 0.645	& 95.6	& 0.012	& 0.608	& 0.608	& 92.6   \\
				\midrule
				exp & {} & $ \boldsymbol{\hat{\psi}^{t}}$  &  0.003	& 0.068	& 0.068	& 96.8	& -0.001	& 0.049	& 0.049	& 97.4	& 0.000	& 0.424	& 0.424	& 96.8	& -0.003	& 0.297	& 0.297	& 94.8   \\
				{} & (0.5, 0.5) & $  \boldsymbol{\hat{\psi}^{n}} $ & -0.212	& 0.250	& 0.328	& 84.8	& 0.196	& 0.274	& 0.337	& 87.4	& -0.184	& 0.424	& 0.462	& 91.8	& 0.191	& 0.356	& 0.404	& 91.6  \\
				{} & {} & $ \boldsymbol{\hat{\psi}^{nb}}  $  & -0.112	& 0.199	& 0.229	& 91.4	& 0.115	& 0.222	& 0.250	& 90.4	& -0.080	& 0.426	& 0.433	& 93.0	& 0.090	& 0.341	& 0.353	& 91.2   \\
				{} & {} & $ \boldsymbol{\hat{\psi}^{rc}} $  & -0.005	& 0.214	& 0.214	& 95.4	& 0.006	& 0.243	& 0.244	& 94.8	& 0.026	& 0.453	& 0.454	& 95.8	& -0.015	& 0.379	& 0.379	& 92.2  \\
				{} & (0.7, 0.7) & $ \boldsymbol{\hat{\psi}^{n}} $ & -0.355	& 0.292	& 0.459	& 74.6	& 0.347	& 0.302	& 0.460	& 76.6	& -0.304	& 0.414	& 0.514	& 89.2	& 0.310	& 0.358	& 0.473	& 84.0  \\
				{} & {} & $ \boldsymbol{\hat{\psi}^{nb}}  $  & -0.207	& 0.247	& 0.323	& 84.8	& 0.191	& 0.272	& 0.332	& 87.8	& -0.155	& 0.425	& 0.452	& 94.4	& 0.163	& 0.360	& 0.395	& 91.8  \\
				{} & {} & $ \boldsymbol{\hat{\psi}^{rc}} $  & -0.020	& 0.288	& 0.289	& 96.2	& 0.006	& 0.330	& 0.330	& 96.0	& 0.036	& 0.480	& 0.482	& 95.2	& -0.025	& 0.439	& 0.439	& 95.4  \\
				{} & (0.9, 0.9) & $  \boldsymbol{\hat{\psi}^{n}} $ & -0.450	& 0.312	& 0.548	& 68.0	& 0.445	& 0.302	& 0.538	& 65.2	& -0.460	& 0.404	& 0.612	& 79.4	& 0.448	& 0.340	& 0.562	& 71.0  \\
				{} & {} & $ \boldsymbol{\hat{\psi}^{nb}}  $  & -0.269	& 0.278	& 0.387	& 82.6	& 0.265	& 0.297	& 0.398	& 83.4	& -0.280	& 0.424	& 0.508	& 90.6	& 0.275	& 0.364	& 0.456	& 87.0  \\
				{} & {} & $ \boldsymbol{\hat{\psi}^{rc}} $  & 0.009	& 0.352	& 0.353	& 96.4	& -0.012	& 0.402	& 0.402	& 94.2	& -0.003	& 0.517	& 0.517	& 95.8	& -0.001	& 0.496	& 0.496	& 95.0  \\
				\midrule
				com & {} & $ \boldsymbol{\hat{\psi}^{t}}$  &  0.003	& 0.070	& 0.070	& 97.6	& -0.003	& 0.051	& 0.051	& 97.4	& -0.003	& 0.104	& 0.104	& 95.8	& 0.002	& 0.074	& 0.074	& 96.6    \\
				{} & (0.5, 0.5) & $  \boldsymbol{\hat{\psi}^{n}}  $ & -0.198	& 0.083	& 0.215	& 33.2	& 0.197	& 0.062	& 0.207	& 11.4	& -0.203	& 0.103	& 0.228	& 49.8	& 0.203	& 0.075	& 0.216	& 23.2  \\
				{} & {} & $ \boldsymbol{\hat{\psi}^{nb}}  $  &  -0.101	& 0.081	& 0.129	& 75.0	& 0.101	& 0.064	& 0.119	& 66.2	& -0.107	& 0.104	& 0.149	& 83.8	& 0.106	& 0.078	& 0.131	& 74.4  \\
				{} & {} & $ \boldsymbol{\hat{\psi}^{rc}} $  & 0.004	& 0.085	& 0.085	& 96.0	& -0.004	& 0.069	& 0.069	& 96.0	& -0.002	& 0.110	& 0.110	& 94.6	& 0.002	& 0.086	& 0.086	& 95.6   \\
				{} & (0.7, 0.7) & $  \boldsymbol{\hat{\psi}^{n}}  $ &  -0.334	& 0.083	& 0.345	& 2.6	& 0.331	& 0.058	& 0.337	& 0.0	& -0.329	& 0.102	& 0.345	& 9.8	& 0.328	& 0.069	& 0.335	& 0.4  \\
				{} & {} & $ \boldsymbol{\hat{\psi}^{nb}}  $  &  -0.190	& 0.083	& 0.207	& 39.2	& 0.187	& 0.063	& 0.197	& 17.4	& -0.186	& 0.104	& 0.213	& 56.2	& 0.185	& 0.076	& 0.200	& 29.4  \\
				{} & {} & $ \boldsymbol{\hat{\psi}^{rc}} $  &  -0.003	& 0.092	& 0.092	& 93.6	& 0.000	& 0.076	& 0.076	& 96.0	& 0.002	& 0.115	& 0.116	& 95.4	& -0.002	& 0.092	& 0.092	& 95.0   \\
				{} & (0.9, 0.9) & $ \boldsymbol{\hat{\psi}^{n}} $ & -0.448	& 0.082	& 0.456	& 0.0	& 0.448	& 0.053	& 0.451	& 0.0	& -0.450	& 0.100	& 0.461	& 0.8	& 0.447	& 0.063	& 0.451	& 0.0  \\
				{} & {} & $ \boldsymbol{\hat{\psi}^{nb}} $  &  -0.275	& 0.083	& 0.288	& 10.0	& 0.275	& 0.060	& 0.281	& 0.2	& -0.278	& 0.103	& 0.297	& 23.2	& 0.275	& 0.072	& 0.284	& 2.6   \\
				{} & {} & $ \boldsymbol{\hat{\psi}^{rc}} $  &  -0.002	& 0.098	& 0.098	& 94.4	& 0.000	& 0.082	& 0.082	& 94.2	& -0.004	& 0.121	& 0.121	& 93.6	& 0.000	& 0.098	& 0.098	& 95.2 \\
				\midrule
			\end{tabular}
		\end{adjustbox}
	\end{sidewaystable}

	\begin{sidewaystable}[hbt!]
		\caption{Prediction accuracy of optimal DTR}
		\label{tab:predoptDTR}
		\centering
		\begin{tabular}{ccrrrrrrrrrrrrrrrrrr}
			\midrule
			{} & {} & \multicolumn{6}{c}{Stage 2 (\%)} & \multicolumn{6}{c}{Stage 1 (\%)}  & \multicolumn{6}{c}{Stage 2 \& Stage 1 (\%)}  \\
			\cmidrule{3-8} \cmidrule{9-14} \cmidrule{15-20}  
			$\sigma_{2}$ & $\sigma_{1}$ & \multicolumn{1}{c}{nt} & \multicolumn{1}{c}{nbt} & \multicolumn{1}{c}{ct} & \multicolumn{1}{c}{nn} & \multicolumn{1}{c}{nbnb} & \multicolumn{1}{c}{cc} & \multicolumn{1}{c}{nt} & \multicolumn{1}{c}{nbt} & \multicolumn{1}{c}{ct} & \multicolumn{1}{c}{nn} & \multicolumn{1}{c}{nbnb} & \multicolumn{1}{c}{cc} & \multicolumn{1}{c}{nt} & \multicolumn{1}{c}{nbt} & \multicolumn{1}{c}{ct} & \multicolumn{1}{c}{nn} & \multicolumn{1}{c}{nbnb} & \multicolumn{1}{c}{cc} \\
			\midrule
			0.5 &  0.5  &  95.6	& 97.5	& 98.4	& 87.0	& 90.7	& 90.7	& 95.4	& 97.0	& 97.6	& 86.9	& 90.5	& 90.5	& 91.2	& 94.6	& 96.1	& 75.6	& 82.1	& 82.1  \\
			{}   &  0.7  &  95.5	& 97.5	& 98.4	& 87.0	& 90.7	& 90.7	& 92.0	& 95.7	& 97.6	& 83.0	& 87.3	& 87.3	& 87.9	& 93.3	& 96.0	& 72.2	& 79.2	& 79.2   \\
			{}   &  0.9  &  95.5	& 97.5	& 98.3	& 87.0	& 90.7	& 90.7	& 87.6	& 93.6	& 97.5	& 79.8	& 84.5	& 84.5	& 83.7	& 91.3	& 95.9	& 69.4	& 76.7	& 76.7  \\
			0.7&  0.5  & 91.8	& 95.9	& 98.3	& 83.1	& 87.5	& 87.5	& 95.4	& 97.0	& 97.6	& 86.9	& 90.5	& 90.5	& 87.6	& 93.0	& 96.0	& 72.2	& 79.2	& 79.2  \\
			{}   &  0.7 & 91.7	& 95.8	& 98.3	& 83.0	& 87.5	& 87.5	& 92.0	& 95.7	& 97.8	& 82.9	& 87.4	& 87.4	& 84.4	& 91.7	& 96.1	& 68.9	& 76.4	& 76.4  \\
			{}   &  0.9  &  92.0	& 96.0	& 98.4	& 83.1	& 87.5	& 87.5	& 87.4	& 93.5	& 97.5	& 79.8	& 84.5	& 84.5	& 80.4	& 89.7	& 96.0	& 66.2	& 73.9	& 73.9   \\
			0.9 & 0.5  & 87.4	& 93.7	& 98.5	& 79.9	& 84.7	& 84.7	& 95.3	& 97.0	& 97.7	& 86.9	& 90.5	& 90.5	& 83.3	& 90.9	& 96.2	& 69.4	& 76.7	& 76.7  \\
			{}  &  0.7  & 87.3	& 93.6	& 98.3	& 79.9	& 84.7	& 84.7	& 91.9	& 95.6	& 97.7	& 83.0	& 87.4	& 87.4	& 80.2	& 89.5	& 96.0	& 66.3	& 74.0	& 74.0   \\
			{}  &  0.9  & 87.4	& 93.7	& 98.4	& 79.9	& 84.7	& 84.7	& 87.6	& 93.6	& 97.6	& 79.8	& 84.5	& 84.5	& 76.6	& 87.7	& 96.0	& 63.8	& 71.6	& 71.6 \\
			\midrule
		\end{tabular}
		\begin{tablenotes}[flushleft]
			\item Note: opt: true optimal DTR, nt: optimal DTR estimated using $\boldsymbol{\hat{\psi}^{n}}$ and true covariate, nbt: optimal DTR estimated using $\boldsymbol{\hat{\psi}^{nb}}$ and true covariate, ct: optimal DTR estimated using $\boldsymbol{\hat{\psi}^{rc}}$ and  true covariate, nn: optimal DTR estimated using $\boldsymbol{\hat{\psi}^{n}}$ and a single surrogate, nbnb: optimal DTR estimated using $\boldsymbol{\hat{\psi}^{nb}}$ and averaged surrogate, cc: optimal DTR estimated using $\boldsymbol{\hat{\psi}^{rc}}$ and regression calibration estimates
		\end{tablenotes}
	\end{sidewaystable}

\begin{sidewaystable}[hbt!]
		\caption{Predicted optimal value function (standard deviations)}
		\label{tab:pred.vfun}
		\centering
		\begin{tabular}{ccrrrrrrr}
			\midrule
			$\sigma_{2}$ & $\sigma_{1}$ & \multicolumn{1}{c}{opt} & \multicolumn{1}{c}{nt} & \multicolumn{1}{c}{nbt} & \multicolumn{1}{c}{ct} & \multicolumn{1}{c}{nn} & \multicolumn{1}{c}{nbnb} & \multicolumn{1}{c}{cc} \\
			\midrule
			0.5 &  0.5  &  3.392  (0.029)& 3.258	(0.051)	& 3.322	(0.052)	& 3.395	(0.055)	& 3.321	(0.054)	& 3.357	(0.053)	& 3.357	(0.054)	  \\
			{}   &  0.7  & 3.392	(0.029)	& 3.217	(0.051)	& 3.295	(0.051)	& 3.395	(0.056)	& 3.296	(0.054)	& 3.342	(0.053)	& 3.342	(0.055)   \\
			{}   &  0.9 & 3.392 (0.029)	& 3.184	(0.051)	& 3.266	(0.051)	& 3.395	(0.058)	& 3.272	(0.055)	& 3.325	(0.053)	& 3.326	(0.056)  \\
			\midrule
			0.7 &  0.5  & 3.392	 (0.029)	& 3.217	(0.052)	& 3.295	(0.052)	& 3.395	(0.056)	& 3.296	(0.054)	& 3.342	(0.053)	& 3.342	(0.055)  \\
			{}   &  0.7  & 3.394	 (0.029)	& 3.179	(0.051)	& 3.269	(0.050)	& 3.396	(0.055)	& 3.272	(0.053)	& 3.328	(0.052)	& 3.329	(0.054)  \\
			{}   &  0.9  &  3.396	 (0.028)	& 3.147	(0.050)	& 3.241	(0.047)	& 3.396	(0.053)	& 3.251	(0.052)	& 3.312	(0.049)	& 3.312	(0.051)  \\
			\midrule
			0.9 & 0.5  & 3.397	 (0.028)	& 3.186	(0.052)	& 3.269	(0.052)	& 3.398	(0.056)	& 3.273	(0.055)	& 3.327	(0.054)	& 3.328	(0.056)  \\
			{}  &  0.7 &  3.396	(0.029)	& 3.144	(0.054)	& 3.239	(0.052)	& 3.394	(0.059)	& 3.247	(0.057)	& 3.310	(0.055)	& 3.310	(0.057)  \\
			{}  &  0.9  & 3.397	 (0.028)	& 3.114	(0.051) & 3.214	(0.048)	& 3.398	(0.055)	& 3.227	(0.054)	& 3.296	(0.051)	& 3.297	(0.053)  \\
			\midrule
		\end{tabular}
		\begin{tablenotes}[flushleft]
			\item Note: opt: true optimal DTR, nt: optimal DTR estimated using $\boldsymbol{\hat{\psi}^{n}}$ and true covariate, nbt: optimal DTR estimated \\
			using $\boldsymbol{\hat{\psi}^{nb}}$ and true covariate, ct: optimal DTR estimated using $\boldsymbol{\hat{\psi}^{rc}}$ and  true covariate, nn: optimal DTR estimated \\
			using $\boldsymbol{\hat{\psi}^{n}}$ and a single surrogate, nbnb: optimal DTR estimated using $\boldsymbol{\hat{\psi}^{nb}}$ and averaged surrogate, cc: optimal DTR \\
			estimated using $\boldsymbol{\hat{\psi}^{rc}}$ and regression calibration estimates
		\end{tablenotes}
	\end{sidewaystable}

\begin{sidewaystable}[hbt!]
		\begin{center}
			\begin{minipage}{\textwidth}
				\caption{Analysis results of the STAR*D data for the blip parameters}\label{tab:stard}
				\begin{tabular}{lrrrrrrrrr}
					\midrule
					{} & \multicolumn{3}{@{}c@{}}{Clinicians} & \multicolumn{3}{@{}c@{}}{Patients} & \multicolumn{3}{@{}c@{}}{Corrected} \\
					\cmidrule{2-4}\cmidrule{5-7} \cmidrule{8-10} 
					Variables & \multicolumn{1}{c}{Est} & \multicolumn{1}{c}{SE} & \multicolumn{1}{c}{95\%CI} & \multicolumn{1}{c}{Est} & \multicolumn{1}{c}{SE} & \multicolumn{1}{c}{95\%CI} & \multicolumn{1}{c}{Est} & \multicolumn{1}{c}{SE} & \multicolumn{1}{c}{95\%CI}  \\
					\midrule
					A$_{2}$ & -0.330  & 0.455 & (-1.221, 0.561) &  -0.436   & 0.367 & (-1.155, 0.284) & -0.351   & 0.396 & (-1.128, 0.426)  \\ 
					A$_{2}$P$_{2}$ & 0.243  &  0.436 & (-0.611, 1.098)  & 0.620   & 0.252  & (0.127, 1.113) & 0.404   & 0.271 & (-0.128, 0.935)  \\ 
					A$_{2}$S$_{2}$ & -0.205  &  1.260 & (-2.674, 2.264)  & -1.011   & 0.979 & (-2.931, 0.908) & -0.539   & 1.316 & (-3.119, 2.041)   \\ 
					A$_{2}$Q$_{2}$ & 0.441 &   0.765 & (-1.060, 1.941)  & 0.936   & 0.459  & (0.036, 1.836) & 0.702   & 0.644 & (-0.561, 1.965)  \\
					\midrule
					A$_{1}$ & -0.194 &   0.159 & (-0.506, 0.117)  & -0.084   & 0.167 & (-0.411, 0.244)  & -0.150  &  0.157 & (-0.458, 0.159)  \\
					A$_{1}$P$_{1}$ & 0.186 &  0.183 & (-0.173, 0.545) & 0.089   & 0.187 & (-0.278, 0.457) & 0.148   & 0.178 & (-0.201, 0.497)  \\
					A$_{1}$S$_{1}$ & 0.062  &  0.091 & (-0.116, 0.241) & 0.069   & 0.093 & (-0.114, 0.252) & 0.090   & 0.091 & (-0.088, 0.268)  \\
					A$_{1}$Q$_{1}$ & 0.048  & 0.066 & (-0.081, 0.178)  & -0.064   & 0.084 & (-0.229, 0.101)  & -0.014   & 0.070 & (-0.152, 0.124)  \\
					\midrule
				\end{tabular}
				\begin{tablenotes}[flushleft]
					\item Note: Est: estimates, SE: standard error, CI: confidence interval
				\end{tablenotes}
			\end{minipage}
		\end{center}
	\end{sidewaystable}
	

\end{document}